\documentclass[conference]{IEEEtran}
\usepackage[dvipdfmx]{graphicx}
\usepackage{epsfig}
\ifCLASSINFOpdf
\usepackage[fleqn]{amsmath}
\usepackage[dvipdfm]{graphicx}
\usepackage{latexsym}
\usepackage[varg]{txfonts}
\usepackage{bm}
\usepackage{color}
\usepackage{cases}
\usepackage{amsmath,amssymb}
\usepackage{subfig}
\usepackage{subfigure}

\else
\fi
\usepackage[cmex10]{amsmath}
\begin{document}
\title{Bilateral Control of Two Finger Joints\\ Using Functional Electrical Stimulation}

%\author{\IEEEauthorblockN{Yuu Hasegawa, Tomoya Kitamura, Sho Sakaino, Toshiaki Tsuji}
%\IEEEauthorblockA{Graduate School of Science and Engineering, \\Saitama University,\\
%255, Shimo-ohkubo, Sakura-ku, Saitama, 338--8570, Japan\\
%Email: y.hasegawa.470@ms.saitama-u.ac.jp, t.kitamura.360@ms.saitama-u.ac.jp, \\
%sakaino@mail.saitama-u.ac.jp, tsuji@ees.saitama-u.ac.jp\\}
%Email: http://www.michaelshell.org/contact.html}
%}

\author{\IEEEauthorblockN{Yuu Hasegawa}
\IEEEauthorblockA{Graduate School of Science and Engineering, \\Saitama University,\\
255, Shimo-ohkubo, Sakura-ku, Saitama, 338--8570, Japan\\
Email: y.hasegawa.470@ms.saitama-u.ac.jp\\}
%Email: http://www.michaelshell.org/contact.html}
\and
\IEEEauthorblockN{Tomoya Kitamura}
\IEEEauthorblockA{Graduate School of Science and Engineering, \\Saitama University,\\
255, Shimo-ohkubo, Sakura-ku, Saitama, 338--8570, Japan\\
Email: t.kitamura.360@ms.saitama-u.ac.jp\\}
%Email: http://www.michaelshell.org/contact.html}
\and
\IEEEauthorblockN{Sho Sakaino}
\IEEEauthorblockA{Graduate School of Science and Engineering, \\Saitama University,\\
255, Shimo-ohkubo, Sakura-ku, Saitama, 338--8570, Japan\\
JST PRESTO\\
Email: sakaino@mail.saitama-u.ac.jp\\}
\and
\IEEEauthorblockN{Toshiaki Tsuji}
\IEEEauthorblockA{Graduate School of \\Science and Engineering, Saitama University,\\
255, Shimo-ohkubo, Sakura-ku, Saitama, 338--8570, Japan\\
Email: tsuji@ees.saitama-u.ac.jp\\}
}

%Email: http://www.michaelshell.org/contact.html}

%\and
%\IEEEauthorblockN{Sho Sakaino and Toshiaki Tsuji}
%\IEEEauthorblockA{The University of Saitama\\
%255, Shimo-ohkubo, Sakura-ku, Saitama, 338--8570, Japan\\
%Fax: (888) 555--1212\\
%Email: sakaino@mail.saitama-u.ac.jp, tsuji@ees.saitama-u.ac.jp}}


%Telephone: (800) 555--1212, Fax: (888) 555--1212}

%IEEEspecialpapernotice{(Invited Paper)}

\maketitle

\begin{abstract}
Bilateral control, a remote-control technique, is used to work at a distance. However, many existing bilateral control systems have two common problems: 1) it is difficult to create a system like a human hand,  that has multiple degrees of freedom and 2) if the mechanism becomes too complicated, operators feel restrained and experience discomfort. Because, for these reasons, the bilateral control of fingers has not been accomplished to date, we aimed to overcome this by applying functional electrical stimulation~(FES). In  our experiments, through an adhesive electrode pad, electrical stimulation was delivered to the muscles that flex and expand the metacarpophalangeal joints of the thumb and middle finger. Position-symmetrical bilateral control was implemented so that the deviation of the master's and slave's positions relative to each other was zero degrees. A sliding mode controller was used as a position controller. 
We found it possible to control multiple degrees of freedom; however, we found areas where the number of tracking errors was large. We speculated that the middle finger did not bend, because the arm rotates as the thumb was abduction, therefore the position of the motor point of the middle finger deviates from the position of the pad.

%\boldmath
%This paper describes the reduction of chattering in functional electrical stimulation (FES) with Smith compensator. We proposed a bilateral controller using FES. In the bilateral controller, in addition to normal remote control, the reaction forces of the distant location are transmitted back to the operator. Therefore, the operator feels like being in the distant location. In controlling of the human body using FES, a dead time occurs. As a result, high-frequency vibration (called chattering) occurs. In this paper, we propose a control method suppressing chattering in FES. In the experiment, the performance of the proposed method was verified by target value control and bilateral control. In the target value control, chattering was suppressed by the proposed method. However, in the bilateral control, chattering was not be suppressed by the interference of the time delay of the two subjects.\\
\end{abstract}



% \centering \bfseries EDICS Category: 3-BBND 
% \fi
\IEEEpeerreviewmaketitle

\section{Introduction}
Robotic remote-control technology is useful in many situations, such as extreme environments and the medical field~\cite{c1}~\cite{c2}.

Bilateral control, a type of remote-control technique, can transmit force information between a master and a slave. The force information received in the slave environment can be conveyed to the master~\cite{c3}~\cite{c4}. The operator can experience the actual feeling of the remote environment via bilateral controllers. Conventionally, bilateral control has been researched for use in the medical field and on disaster sites~\cite{c5}~\cite{c6}. 

However, many existing bilateral control systems have two common problems: 1) it is difficult to create a system with multi-degrees of freedom, like the human hand, and 2) as the mechanism becomes complicated, operators feel uncomfortable restrained. In the current paper, we aimed to solve these problems using functional electrical stimulation~(FES) because it has already been demonstrated to achieve bilateral control~\cite{c7}~\cite{c8}; it has been used to restore function, by providing electrical stimulation, to permanently paralyzed limbs resulting from upper motor-neuron disorders such as spinal cord injury and stroke. Research into controlling the human body using FES has been actively conducted since the 1960s~\cite{c9}. In recent years, the use of FES not only for rehabilitation but also to move the body has attracted much attention. 

The following research has been carried on the use of FES for bodily control. Gollee {\it et al.} reproduced standing motions with an intact and a paraplegic subject using linear second Gaussian control of FES~\cite{c10}. Ching {\it et al.} proposed using neural network~(NN) in combination with PID control to improve control performance~\cite{c11}. Ajoudani {\it et al.} proposed control by combining NN and sliding-mode control~\cite{c12}. Tamaki {\it et al.} controlled hands using FES; they showed that it was possible to use hands to help play a musical instrument~\cite{c13}; they used FES to move their fingers and tried to reproduce the object's gripping motion by performing closed-loop control using a force sensor~\cite{c14}. Kitamura {\it et al.} performed bilateral control of an elbow joint using FES~\cite{c15}. These various studies indicate that FES is useful for control. However, the bilateral control of fingers has not yet been reported in any studies to the best of our knowledge. Human fingers are capable of many varied motions; if FES is to be used as a remote-control technology in the future, it is imperative that it can bilaterally control the movement of fingers. Therefore, in this paper, as initial research into the bilateral control of fingers using FES, bilateral control of a movement of the thumb and middle finger was performed by applying electrical stimulation. Moreover, Farhoud {\it et al.} confirmed that control is improved by using a high-order sliding mode control for an operation using FES~\cite{c16}. Therefore, in this paper, bilateral control using FES was also performed using high-order sliding mode control. 

The current paper is organized as follows; Section II describes FES; Section III describes bilateral control, Section IV describes sliding mode control, Section V describes the experimental environment, Section VI describes the experimental method, Section VII describes and discusses the experimental result of the bilateral control, and Section VIII concludes this paper.

\section{Functional Electrical Stimulation}
\begin{figure}[bp]
\centering
\includegraphics[width=40mm]{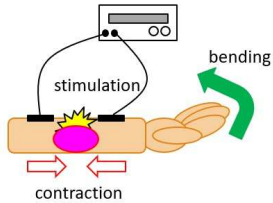}
\caption{Conception diagram of FES}
\label{gainen}
\end{figure}

In this section, FES is described and is conceptually illustrated in Fig.~\ref{gainen}. FES delivers electrical stimulation to peripheral nerves using an external power source and excites the peripheral nerves; in this way, it is possible to drive the body.

\subsection{Stimulation Method}
\begin{figure}[bp]
  \centering
    \begin{tabular}{c}
      % 1
      \begin{minipage}{0.33\hsize}
        \centering
          \includegraphics[width=25mm]{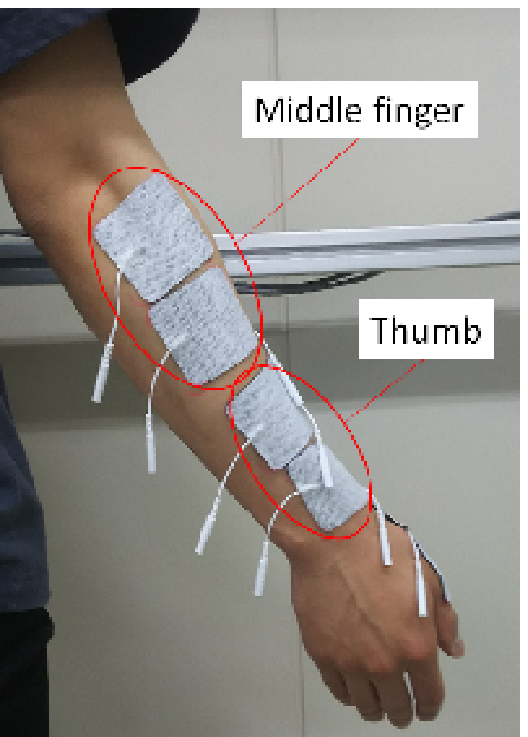}
          \hspace{5mm} [A]Extensor muscle        
      \end{minipage}
\hspace{10mm}
      % 2
      \begin{minipage}{0.33\hsize}
        \centering
          \includegraphics[width=25mm]{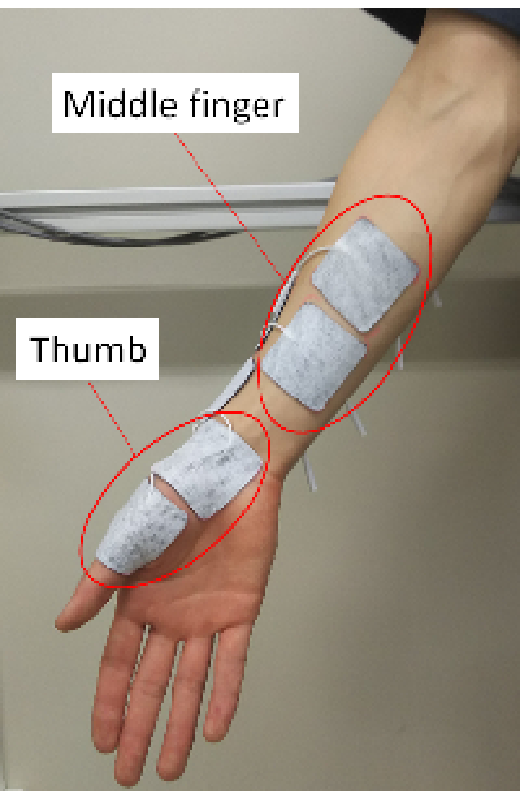}
          \hspace{5mm} [B]Flexor muscle        
      \end{minipage}
    \end{tabular}
    \caption{Stimulus location}
    \label{fig:kinnniku} 
\end{figure}

In this experiment, electrostimulation was delivered to the forearm using an adhesive pad for safety and convenience consideration. To drive the thumb and middle finger, we stimulated the flexor and extensor apollicis brevis muscles, the pollicis brevis muscle, the flexor digitorum superficialis muscle, and the extensor digitorum muscle. The flexor works in the direction that bends the fingers and the extensor works in the direction that extends the fingers. Fig.~\ref{fig:kinnniku} shows the locations of the pads.

%In this section, FES is described. FES is a technique that induces muscle contraction and drives joints by applying electrical stimulation to muscles.

%Here, we describe how to control a human's elbow joint and verify the proposed method. To begin, the subjects' biceps brachii muscle and the triceps brachii muscle were stimulated. When the biceps brachii muscle is stimulated, the elbow joint flexes. Moreover, when the triceps brachii muscle is stimulated, the elbow joint extends. Fig.~\ref{fig:kinnniku} indicates the location of the pads. 

\subsection{Stimulation Waveform}
\begin{figure}[tbp]
\centering
\includegraphics[width=50mm]{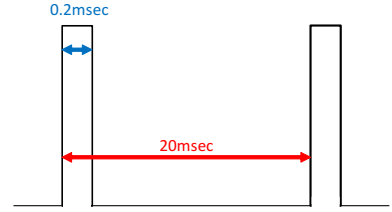}
\caption{Waveform of electrical stimulation}
\label{fig:fes1}
\end{figure}

In this experiment, we used a pulse wave because it is convenient and commonly used. Fig.~\ref{fig:fes1} shows the stimulation waveform with a frequency of 50 Hz and a pulse width of 0.2~msec. The frequency was selected from 20, 50 and 100~Hz the value at which the muscle reacted the most. 
%In this experiment, we use a pulse wave because it is most frequently used conventionally. The stimulating parameters were a pulse width of 0.2 ms and a frequency of 50 Hz. The stimulation waveform is shown in Fig.~\ref{fig:fes1}. In this paper, control was performed by adjusting the voltage amplitude with the maximum set to 45 V. The circuit was configured to conform to the Japanese Industrial Standard (JIS). Therefore, the current flowing through the human body did not exceed 20 mA. 

\section{Bilateral Control}
\begin{figure}[th]
\centering
\includegraphics[width=60mm]{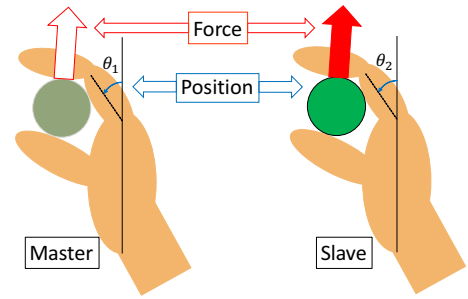}
\caption{Conceptual diagram of bilateral control}
\label{fig:bilate1}
\end{figure}

In this section, bilateral control, a remote-control technique~\cite{c16_1}~\cite{c16_2}, is described and is illustrated in Fig.~\ref{fig:bilate1}. The operator side is called a master and the operated side is called a slave. The slave follows the movement of the master. In addition, forces applied to the slave are feed back to the master. Therefore, the master feels as if it is in the slave's location~\cite{c16_3}~\cite{c16_4}. There are several types of bilateral control; however, in this paper, we used position-symmetrical bilateral control because the control system for this method is simple and stable and does not require a force sensor. 
%Fig.~\ref{block} shows a block diagram of the position symmetrical bilateral control where  subscript $m$ is the master, $s$ is the slave, $\tau^{ref}$ is the reference value of the torque, and $\theta^{res}$ is the angle. Here, the master and slave consist of servo systems. Torque is applied to both the master and slave to make their difference in their relative angles equal to zero. 

%\begin{figure}[htbp]
%\centering
%\includegraphics[width=70mm]{fig/block_bilate3.eps}
%\caption{Block diagram of bilateral control}
%\label{block}
%\end{figure}

%In this section, bilateral control is described. Bilateral control is a type of remote control technology. The operator is called the master and operates the slave. Because bilateral control is remote control, the slave follows the movement of the master. In addition, forces applied to the slave are feedback to the master. Therefore, the master feels like being in the place where the slave is located. In this paper, we use position symmetric bilateral control because
%\begin{itemize}
%\item force sensors are not required,
%\item configuration is easy, and
%\item the control system thus becomes stable.
%\end{itemize}
%For these reasons, we used position symmetric bilateral control. Here, the master and slave consist of servo systems. A force is applied to both the master and slave to make their difference in position zero.

\section{Sliding Mode Control}
In this section, sliding mode control is described. The objective of the sliding mode controller is that the system state vector $x (t)$ converges to the desired state vector $x_d (t)$ depending on the effect of uncertainty such as disturbance or modeling error~\cite{c17}. A general sliding mode control satisfies the control goal when the sliding manifold $s (x, t)$ is on the sliding manifold (ie, when $s (x, t) = 0$). The sliding manifold $s (x, t)$ is defined as follows:
\begin{eqnarray}
s(x, t) &=& (\frac{d}{dx} + \lambda )^{n-1}e(t)       \\
e(t) &=& \theta^{cmd} - \theta^{res}
\end{eqnarray}
where, $\lambda$ is a positive constant. Also, $e(t)$ represents the deflection between the target value and the measured value of the finger joint angle, $\theta^{cmd}$ represents the target value of the angle, and $\theta^{res}$ represents the measured value of the angle. In the current paper, because angle and angular velocity are inputted, the order {\it n} of the input variable is $n = 2$. However, for this classical sliding mode control, the problem of oscillatory, high-frequency input, called chattering, arises; therefore, high order sliding mode~(HOSM) control was used. HOSM control works on higher-order derivatives compared to standard sliding variables. The control target of the HOSM controller can be described by Eq. (3) using the sliding manifold s given in Eq (1). 

\begin{eqnarray}
s = \dot{s} =\ddot{s} = \cdots = s^{r-1} = 0
\end{eqnarray}

In the HOSM controller, the high-order element s, which increases when chattering occurs, consequently converges to zero, and therefore, chattering can be suppressed.

In the current paper, we used a super-twisting algorithm to realize HOSM control~\cite{c17}. This algorithm was developed to avoid the  chattering phenomena. The high-order element s contains up to second order terms that do not depend on the derivative of the sliding variable. In this paper, the control input $u$, satisfying the control target of $s = \dot{s} = 0$, is defined as follows:
\begin{eqnarray}
\begin{cases}
u = -\lambda \times |s|^{\rho} sgn(s) + u_a &\\
\dot{u}_a = -W \times sgn(s) &
\end{cases}
\end{eqnarray}
where, $\lambda, \rho, $ and $W$ are positive constants and $\rho$ is preferably 0.5 when $n = 2$~\cite{c18}. It is difficult to identify the values of $\lambda$ and $W$. Therefore, it was determined that $\lambda = 2.0$ and $W = 0.2$ by trial and error. After the values of these parameters were applied to both subjects, the experiments were carried out.

\section{Experiment Environment}
In this section, the experiment environment is described.

\subsection{Electrical Stimulation Device}
Table~\ref{machine} shows the input and output of electrical stimulation device. We used a maximum output current less than the effective value of 20~mA in accordance with the Japanese Industrial Standard~(JIS) for low-frequency therapy devices. In addition, in consideration of the subject's safety and to reduce their discomfort, we restricted the stimulation voltage to a 30~V maximum.

\begin{table}[t]
\begin{center}
\caption{Input and output of the electrical stimulation device }
\begin{tabular}{|c|c|} \hline
Supply voltage & 50~V  \\ \hline
Output voltage(Effective value) & 45~V \\ \hline
Output current(Effective value) & 20~mA \\ \hline

\end{tabular}
\label{machine}
\end{center}
\end{table}

\subsection{Angle Measuring Device}
\begin{figure}[t]
	\begin{center}
		\begin{tabular}{cc}

		\begin{minipage}{0.5\hsize}
			\begin{center}
			\includegraphics[height=30mm]{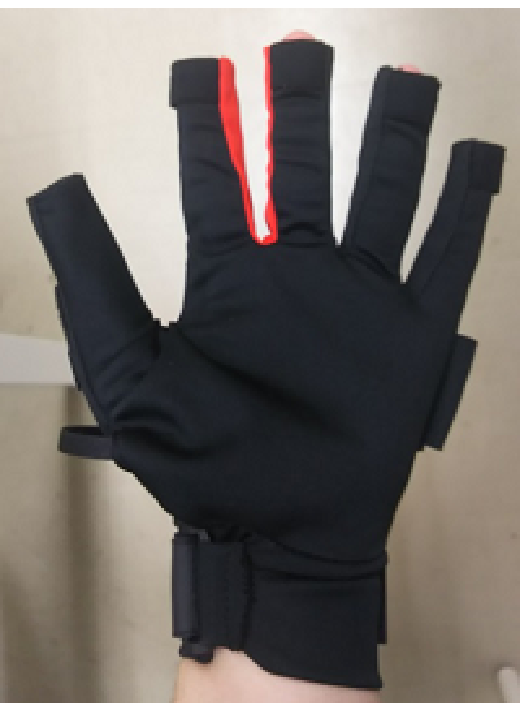}
			\hspace{1.6cm}[A]The palm side
			\end{center}
		\end{minipage}

		\begin{minipage}{0.5\hsize}
			\begin{center}
			\includegraphics[height=30mm]{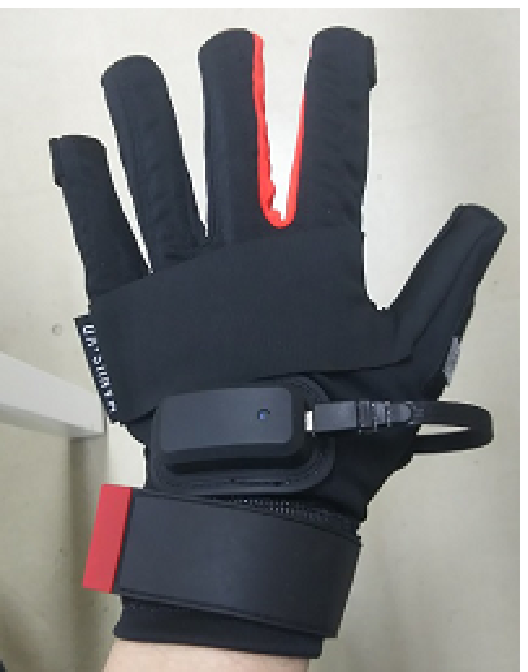}
			\hspace{1.6cm}[B]The back side
			\end{center}
		\end{minipage}

		\end{tabular}
		\caption{Angle measuring device}
		\label{manus}
	\end{center}
\end{figure}

We measured the angle of the finger joint using a data glove (Manus VR, Eindhoven. Netherlands) (Fig.~\ref{manus}). The joint angle of each finger was measured by flexible sensors in the glove's fingers. 

\section{Methods}
\begin{figure}[t]
\centering
\includegraphics[width=35mm]{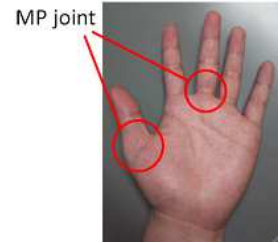}
\caption{MP joint}
\label{MPjoint}
\end{figure}

\begin{figure}[t]
\centering
\includegraphics[width=70mm]{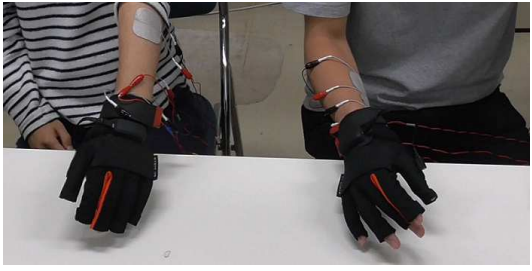}
\caption{Experimental setup}
\label{experiment}
\end{figure}

In this section, the experimental method is described. The subjects, referenced as A and B, were two healthy men aged in their twenties. We explained the content and objectives of the experiment to the subjects and conducted the experiments after obtaining their informed consent. Permission from the ethics committee of the Saitama University was obtained for this experiment. For the experiment, we used the metacarpophalangeal~(MP) joints of the thumb and middle finger, shown in Fig.~\ref{MPjoint}, for bilateral control. The joint angle when the fingers were extended was set to 0~deg and when the fingers were flexed was set to 90~deg. Fig.~\ref{experiment} shows the experimental setup.

\subsection{Determination of Electrical Stimulation Position}
\begin{figure}[t]
\centering
\includegraphics[width=30mm]{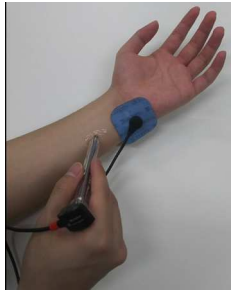}
\caption{Scene of looking for motor point}
\label{mpp}
\end{figure}

For the experiment, we selected a muscle to drive the MP joints of the thumb and middle finger using a motor point pen~(MPP)~(COMPEX). By using an MPP, it was possible to find the part where the muscles tended to respond to electrical stimulation~(motor point)~(Fig.~\ref{mpp}).

\subsection{Control System}
\begin{figure}[t]
\centering
\includegraphics[width=70mm]{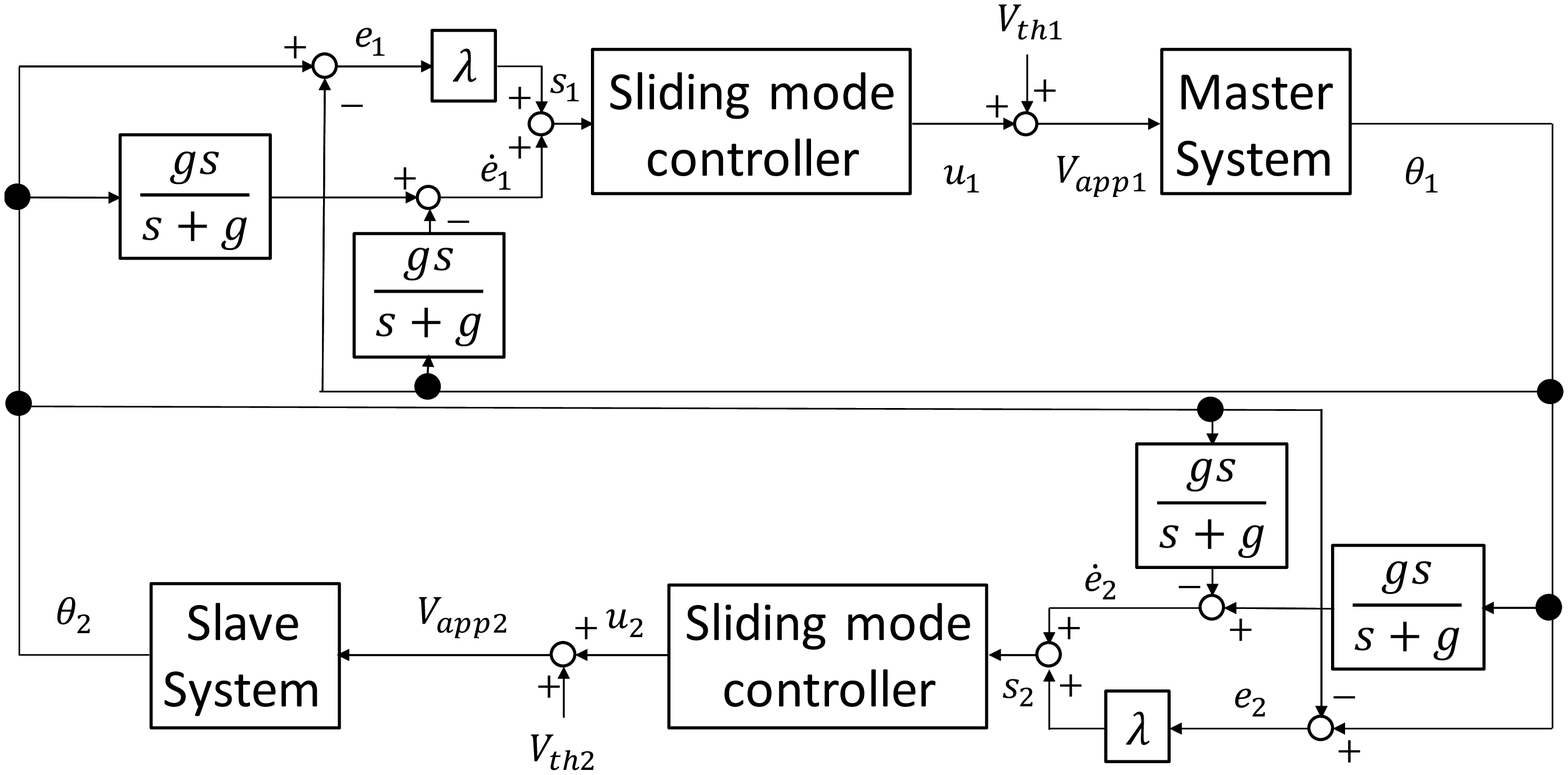}
\caption{Control system}
\label{control}
\end{figure}

The control system used in this experiment is shown in Fig.~\ref{control}. The inputs, $s_1$ and $s_2$, to the sliding mode controller can be expressed by the following equations derived from Eqs (1), (2);
\begin{eqnarray}
s_1 = \dot{\theta}^2 - \dot{\theta}^1 + \lambda(\theta^2 - \theta^1) \\
s_2 = \dot{\theta}^1 - \dot{\theta}^2 + \lambda(\theta^1 - \theta^2)
\end{eqnarray}
where the output $u$ is the stimulation voltage; when $u$ is positive, the muscles which bend the finger joints are stimulated, and when $u$ is negative, the muscles which extend the finger joints are stimulated. However, human muscles do not contract unless a voltage exceeding a threshold voltage is applied. Therefore, in this experiment, the threshold voltage $V_{th}$ was selected for each subject, and the stimulation voltage amplitude $V_{app}$ can be expressed as follows:

\begin{eqnarray}
V_{app} = \begin{cases}
u + V_{th}  &  (u > 0)  \\
0 & (u = 0) \\
-u + V_{th} & (u < 0)           .
\end{cases}
\end{eqnarray}

The angular velocity was obtained by pseudo-differentiating the angle, and the cutoff frequency $g$ was set at 6.28~Hz.

\section{Experimental Results and Discussion}
In this section, we describe the results of the experiment. 

First, two subjects were selected, and one subject became a master and the other subject became a slave. A short instruction gPlease freely move the finger jointh was given to the master. Simultaneously, a short instruction gPlease do not look at the hand of the masterh was given to the slave. The duration of the experiment was 25 s. After this, the master and slave swapped places, and the same 25 s experiment was repeated. 

We conducted three kinds of experiments using FES as follows:
\begin{itemize}
 \item bilateral control of the thumb
 \item bilateral control of the middle finger
 \item bilateral control of both the thumb and the middle finger
\end{itemize}
The experimental results are given below.

\subsection{Bilateral Control of Thumb Using FES}
\begin{figure}[h]
\centering
\includegraphics[width=65mm]{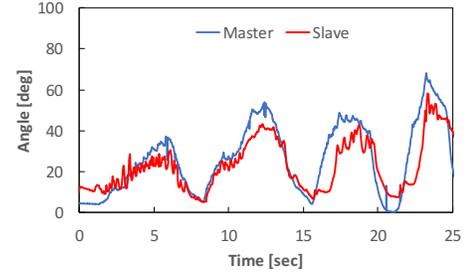}
\caption{Subject A as the master and subject B as the slave}
\label{ha_sa_th}
\end{figure}

\begin{figure}[h]
\centering
\includegraphics[width=65mm]{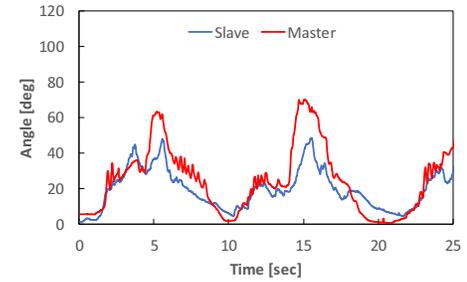}
\caption{Subject B as the master subject A as the slave}
\label{sa_ha_th}
\end{figure}

%First, Table~\ref{parameter} shows threshold voltages $V_{th}$ and frequencies $f$ applied to subjects A and B. 
The results of the experiment using subject A as the master and subject B as the slave are shown in Fig.~\ref{ha_sa_th}. Those with subject A as the slave and subject B as the master are shown in Fig.~\ref{sa_ha_th}. From these result, we confirmed that the thumb of the slave followed the thumb of the master.

%\begin{table}[htbp]
%\begin{center}
%\caption{Parameter of subjects}
%\begin{tabular}{|c||c|c|c|c|} \hline
%\multicolumn{1}{|c||}{} & \multicolumn{2}{c|}{Subject A} & \multicolumn{2}{c|}{Subject B} \\ \hline
%Muscle & $V_{th}$\ [V] & $f$\ [Hz] & $V_{th}$\ [V] & $f$\ [Hz]  \\ \hline
%adductor pollicis muscle & 20 & 100 & 17 & 100\\ \hline
%abductor pollicis longus muscle & 18 & 100 & 19 & 100\\ \hline
%flexor digitorum superficialis muscle & 26 & 50 & 17 & 50\\ \hline
%extensor digitorum muscle & 24 & 50 & 15 & 50\\ \hline

%\end{tabular}
%\label{parameter}
%\end{center}
%\end{table}

\subsection{Bilateral Control of Middle Finger Using FES}
\begin{figure}[h]
\centering
\includegraphics[width=65mm]{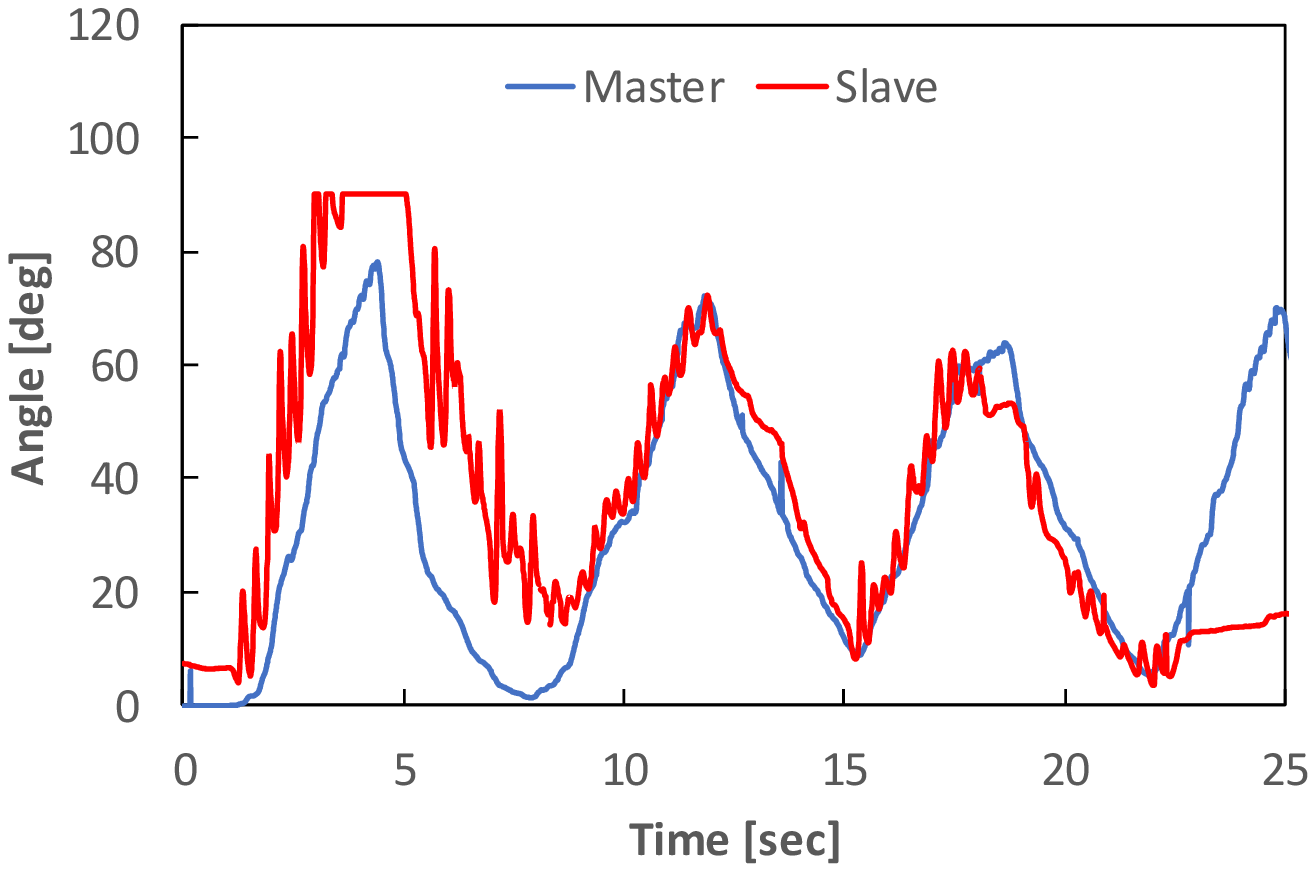}
\caption{Subject A as the master and subject B as the slave}
\label{ha_sa_mi}
\end{figure}

\begin{figure}[h]
\centering
\includegraphics[width=65mm]{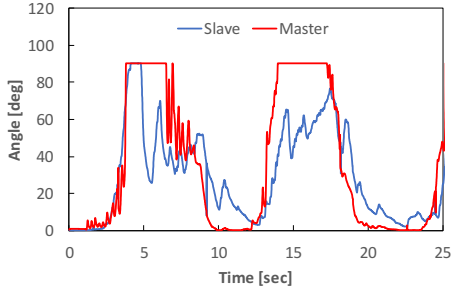}
\caption{Subject B as the master and subject A as the slave}
\label{sa_ha_mi}
\end{figure}

The results of the experiment with subject A as the master and subject B as the slave are shown in Fig.~\ref{ha_sa_mi}. Those with subject A as the slave and subject B as the master are shown in Fig.~\ref{sa_ha_mi}. From these results, it was shown that the movement of the middle finger of the slave generally followed the movement of the middle finger of the master. However, it was evident that there was a period when errors increased.

\subsection{Bilateral Control of Thumb and Middle fingers using FES}
\begin{figure}[h]
\centering
\includegraphics[width=65mm]{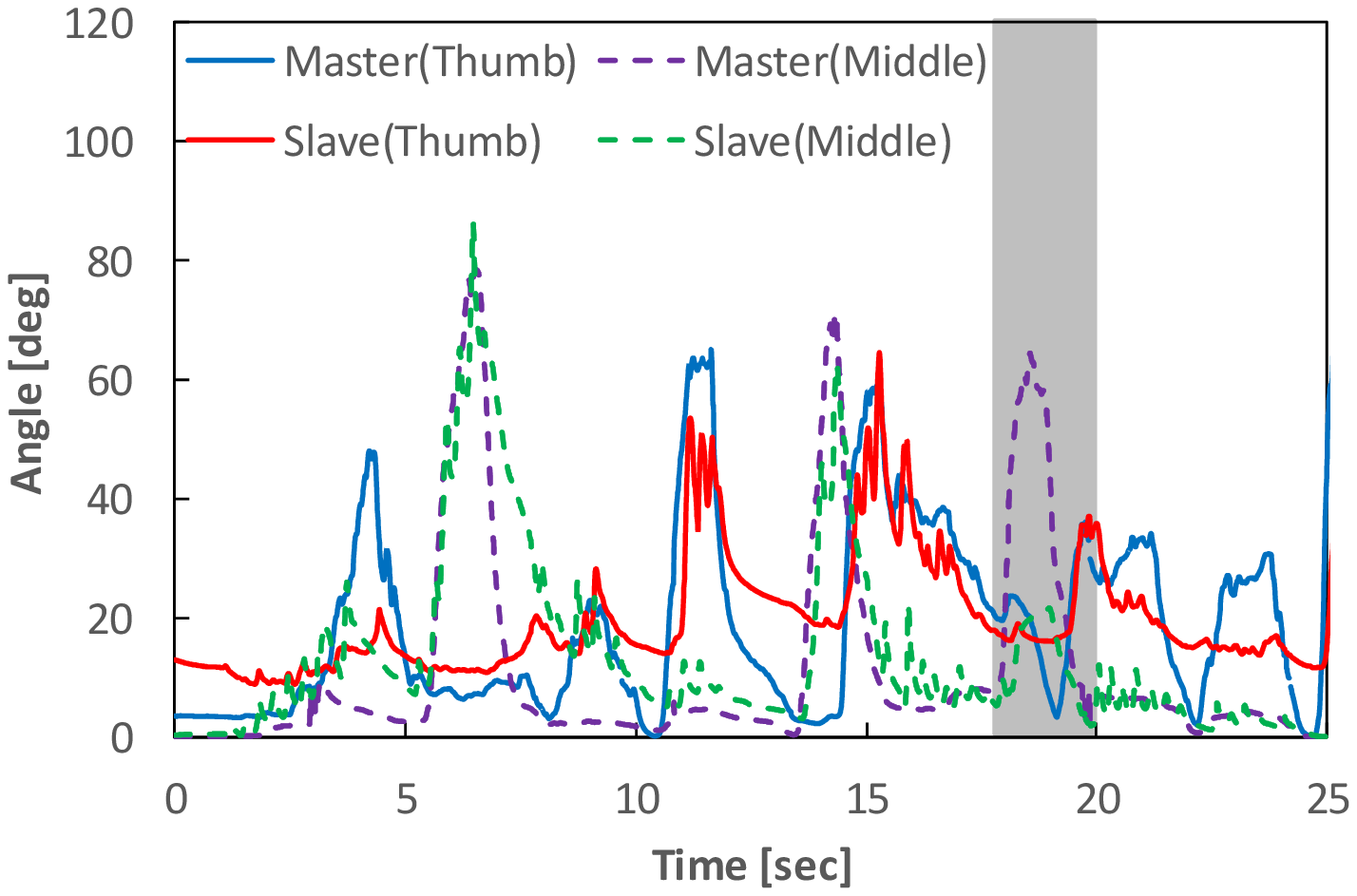}
\caption{Subject A as the master and subject B as the slave}
\label{ha_sa_do}
\end{figure}

\begin{figure}[!h]
\centering
\includegraphics[width=65mm]{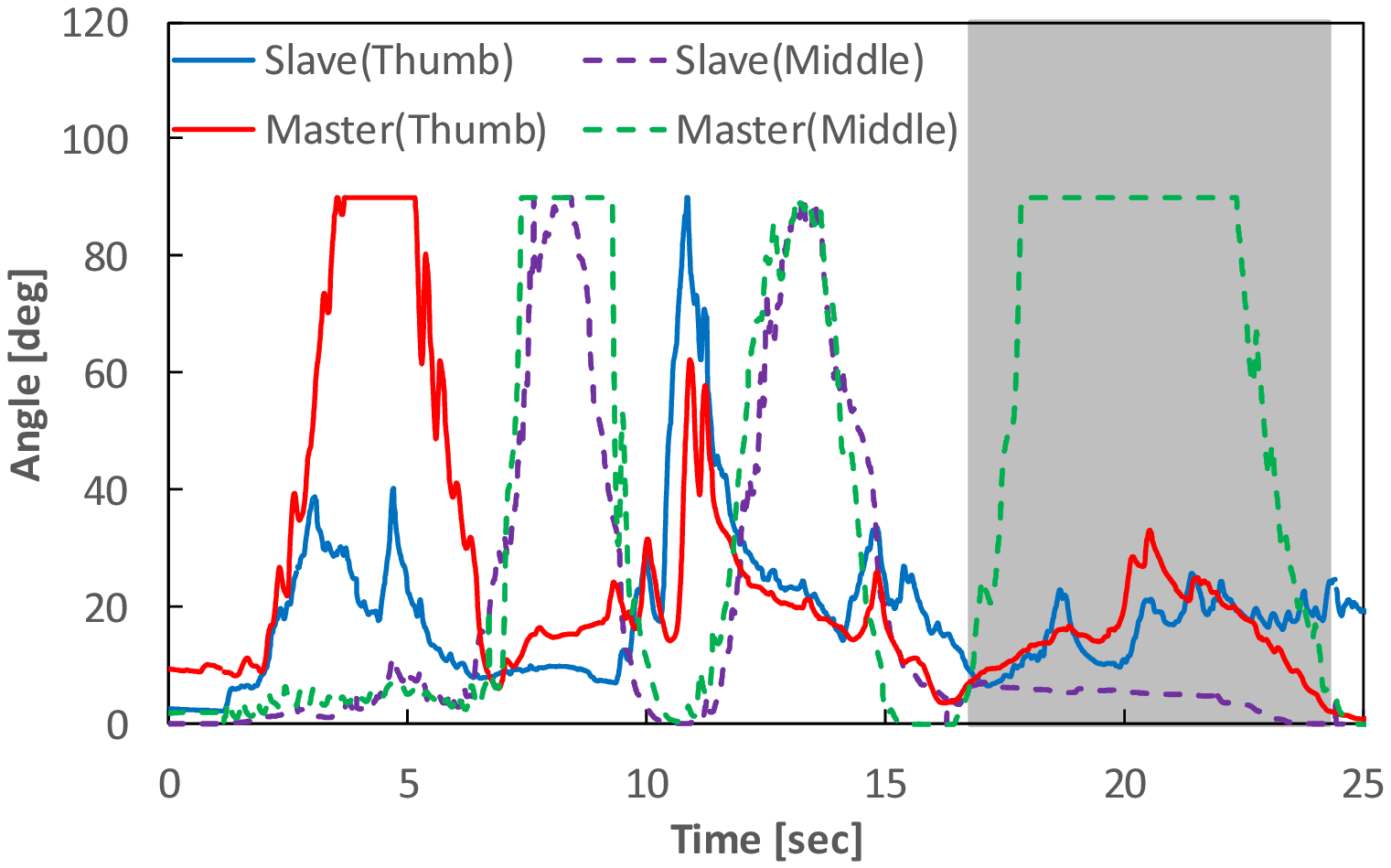}
\caption{Subject B as master and subject A as the slave}
\label{sa_ha_do}
\end{figure}

\begin{figure}[!h]
\centering
\includegraphics[width=65mm]{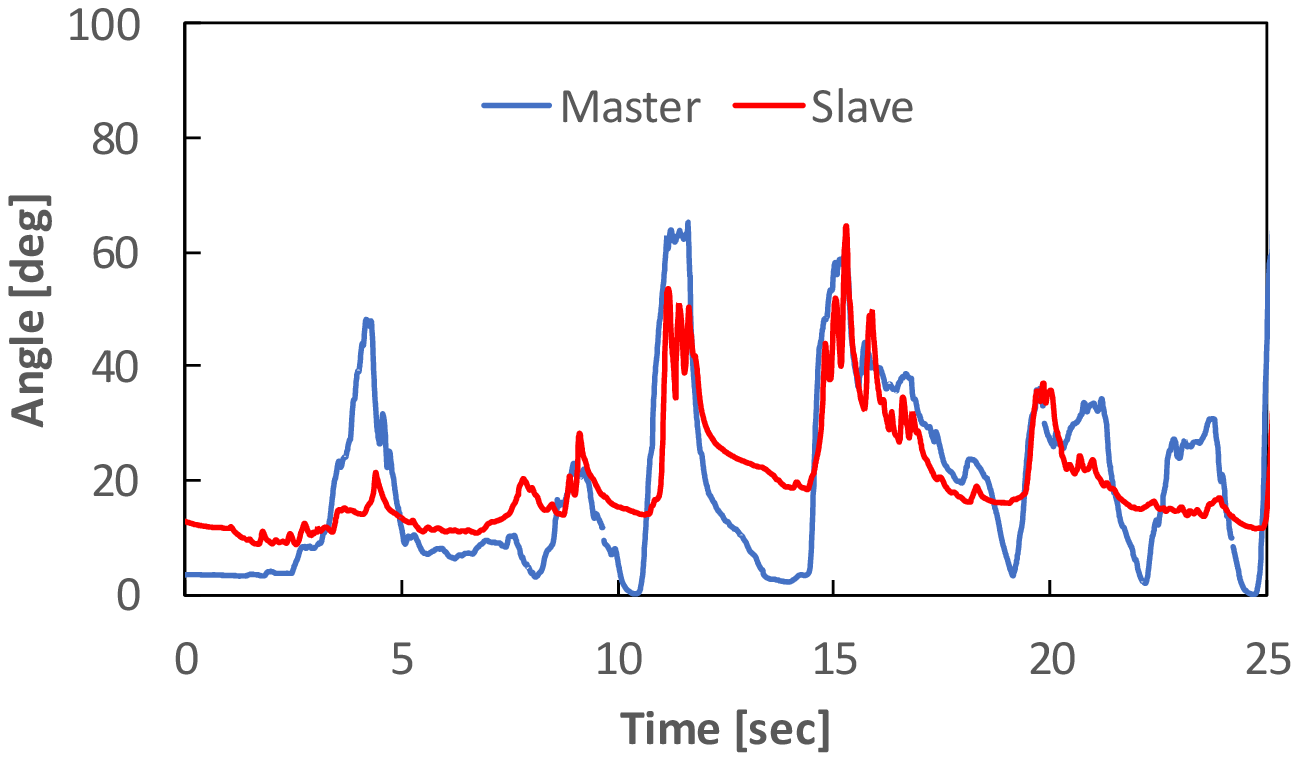}
\caption{Focus on the motion of the thumbs shown in Fig.~\ref{ha_sa_do}}
\label{ha_sa_do_th}
\end{figure}

\begin{figure}[!h]
\centering
\includegraphics[width=65mm]{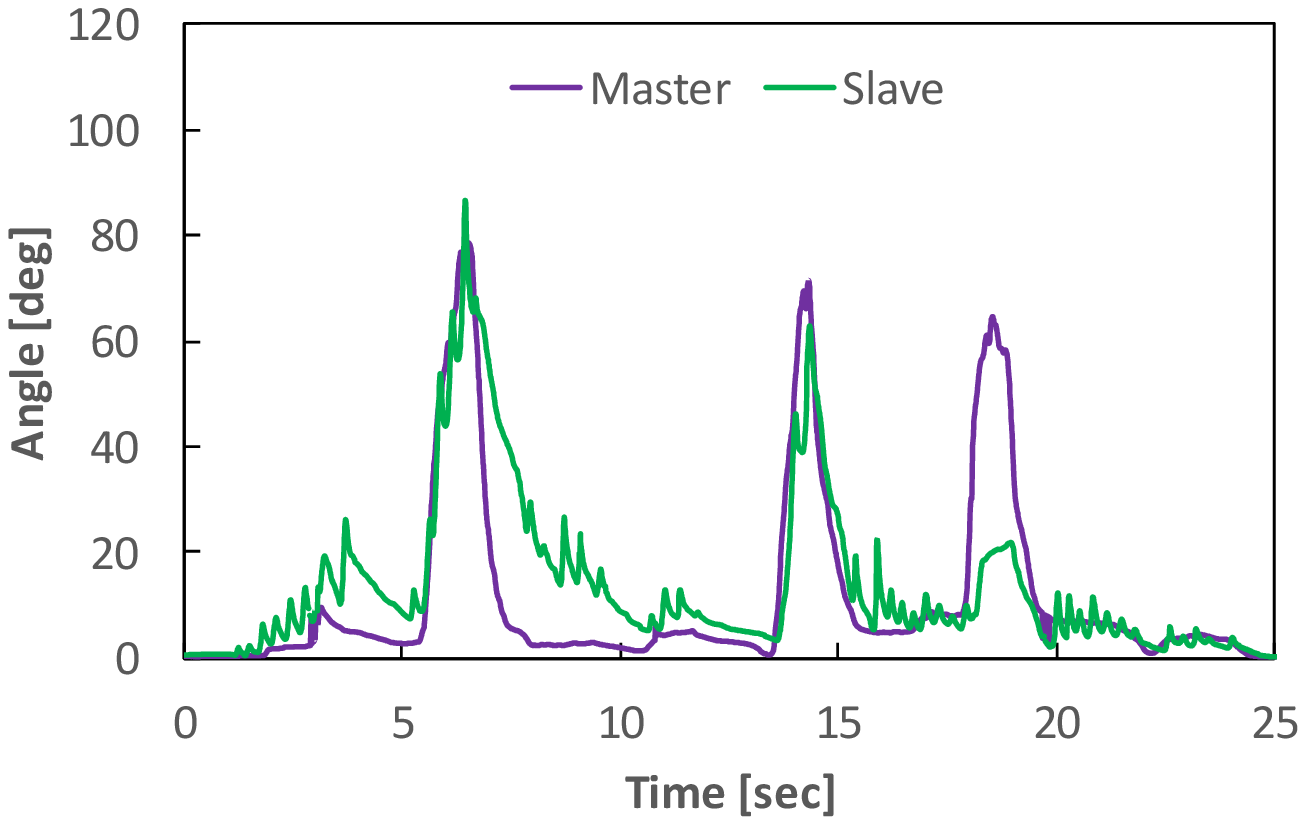}
\caption{Focus on the motion of the middle fingers shown in Fig.~\ref{ha_sa_do}}
\label{ha_sa_do_mi}
\end{figure}

\begin{figure}[!h]
\centering
\includegraphics[width=65mm]{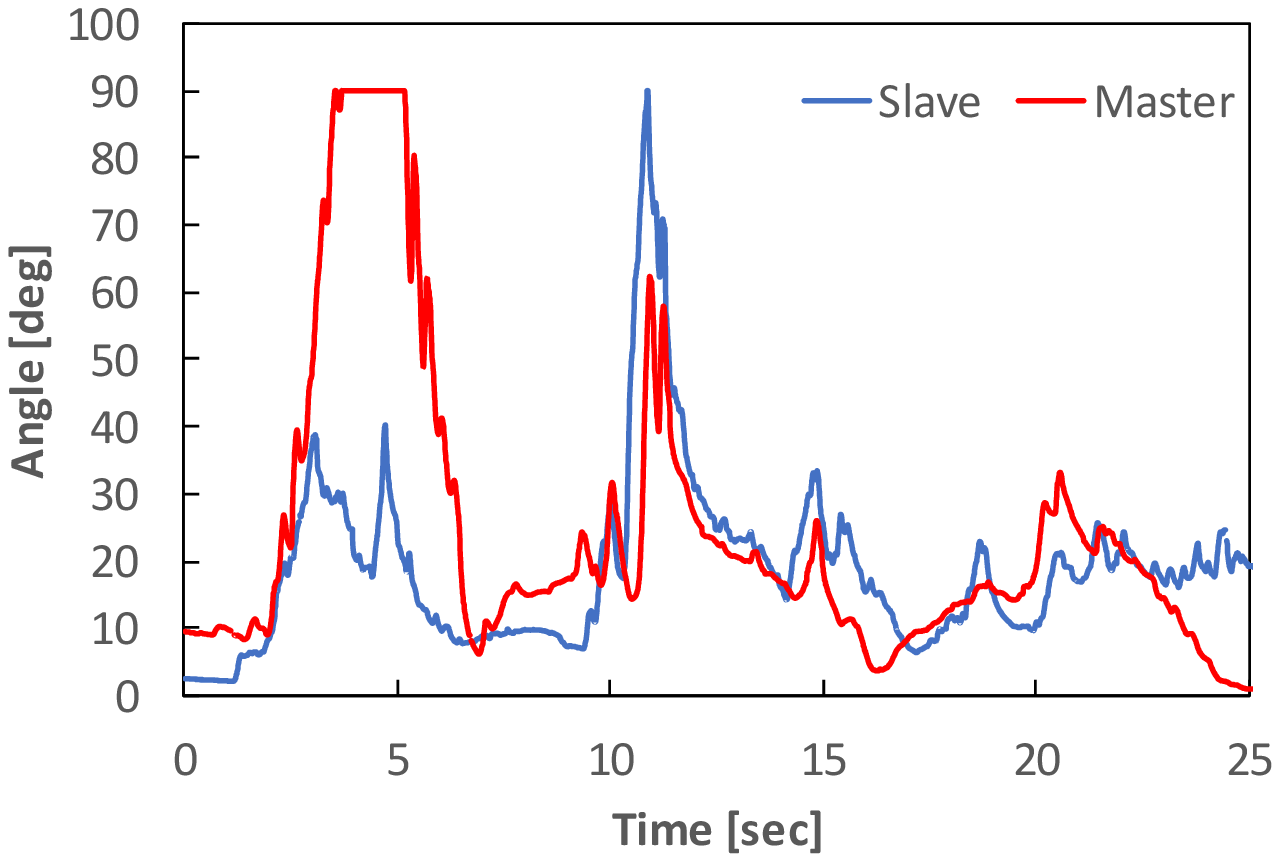}
\caption{Focus on the motion of the thumbs shown in Fig.~\ref{sa_ha_do}}
\label{sa_ha_do_th}
\end{figure}

\begin{figure}[!h]
\centering
\includegraphics[width=65mm]{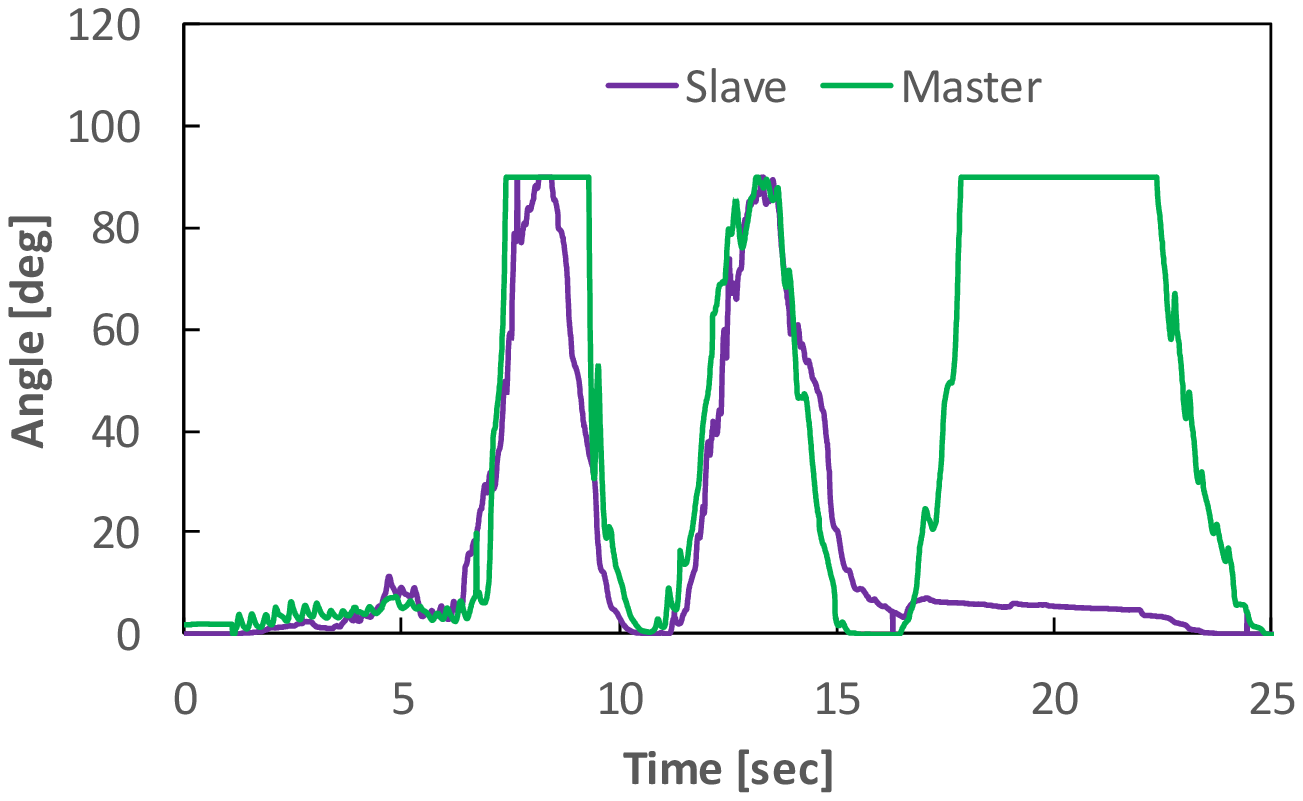}
\caption{Focus on the motion of the middle fingers shown in Fig.~\ref{sa_ha_do}}
\label{sa_ha_do_mi}
\end{figure}

\begin{figure}[!h]
\centering
\includegraphics[width=65mm]{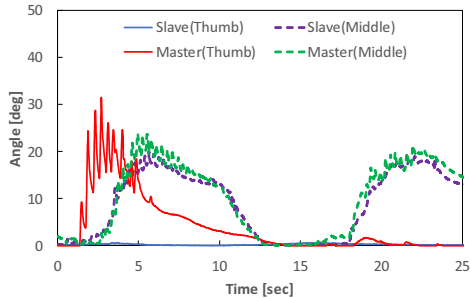}
\caption{Subject B as master and subject A as the slave in contact motion}
\label{sa_ha_hazi}
\end{figure}

The results of the experiments with subject A as the master and subject B as the slave are shown in Fig.~\ref{ha_sa_do}. Those with subject A as the slave and subject B as the master are shown in Fig.~\ref{sa_ha_do}. In addition, Figs.~\ref{ha_sa_do_th}-\ref{sa_ha_do_mi} focus on the motions of the thumb and the middle finger shown in Figs.~\ref{ha_sa_do} and \ref{sa_ha_do}. Figs.~\ref{ha_sa_do}-\ref{sa_ha_do_mi} are free motion data. In addition, contact motion with subject A as the master and subject B as the slave grabbing a cylinder with a diameter of 5~cm  is shown in Fig.~\ref{sa_ha_hazi}. Figs.~\ref{ha_sa_do}-\ref{sa_ha_hazi} also confirmed where the slave followed the motion of the master. However, compared to when the thumb and middle finger were individually, bilaterally controlled, many motions were not followed exactly.

From the gray part of Figs. \ref{ha_sa_do} and \ref{sa_ha_do}, error of the middle finger was increased, when middle fingers were bent for the third time. 
%Therefore, it was inferred that the change in the muscles of the thumb which stimulus was given was affecting the movement of the middle finger. 
We speculated that the middle finger did not bend, because the arm rotated as the thumb was abduction, therefore the position of the motor point of the middle finger deviated from the position of the pad.
%The forearm is arranged so that many muscles overlap. Therefore, when the muscles contract, the position delivering stimulation voltage deviates from the position of the motor point inside the forearm. Therefore it was inferred that the fingers will not move.
%For the above reasons, it was considered that there is a part where the error is large.
In order to solve this problem, we need to investigate the movement of the muscles of the forearm when electric stimulation is delivered, and to devise a mechanism that the stimulation position does not deviate from the motor point. In addition, we speculated that the thumb's time constant became large, because since the setting of the control gain of the thumb could not be set well in Fig.~\ref{sa_ha_hazi}.
%The results suggest that when multiple fingers were controlled simultaneously, interference occurred between the muscles and affected the motion.

%'±'±'©'ç

%'±'±'Ü'Å

%In this paper, we proposed methods that control human bodies using FES with Smith compensator to deal with the dead time peculiar to FES. Experiments were conducted on four subjects. As a result, chattering caused by the dead time was suppressed by the proposed method. On the other hand, chattering was not be suppressed in bilateral control experiments. We speculated that the reason is interference by delay time between the two subjects. In future, we will aim to develop a control system suitable for bilateral control and extend to multiple degrees of freedom.


\section{Conclusion}
We proposed the bilateral control of two degrees of freedom in a thumb and a middle finger using FES. In an experiment, we recorded the motion of the thumbs and the middle fingers of two subjects during bilateral control and confirmed that the slave followed the motion of the master. However, we found a period of time where the tracking error was large. 
We speculated that the middle finger did not bend, because the arm rotated as the thumb was abduction, therefore the position of the motor point of the middle finger deviated from the position of the pad.
%We speculated that this problem was caused by interference between the muscles when the FES was used for multiple muscles simultaneously. 
We will devise a mechanism that the stimulation position does not deviate from the motor point and aim to further increase the degree of freedom.
%We proposed studying stimulation methods that do not cause interference and aim to further increase the degree of freedom.
\section*{Acknowledgement}
This work was entrusted by the KDDI Foundation and JST, PRESTO Grant Number JPMJPR1755, Japan.


%\bibliographystyle{IEEEtran}

\end{document}